\documentclass[letterpaper]{article} 
\usepackage{aaai24}  
\usepackage{times}  
\usepackage{helvet}  
\usepackage{courier}  
\usepackage[hyphens]{url}  
\usepackage{graphicx} 
\usepackage{natbib}  
\usepackage{caption} 
\usepackage{algorithm}
\usepackage{algorithmic}

\usepackage{multirow}
\usepackage{tabularx}
\usepackage{booktabs}
\usepackage{amsmath}
\usepackage{amsthm}
\usepackage{newfloat}
\usepackage{listings}
\DeclareCaptionStyle{ruled}{labelfont=normalfont,labelsep=colon,strut=off} 
\floatstyle{ruled}
\newfloat{listing}{tb}{lst}{}
\floatname{listing}{Listing}
\title{Vector Field Oriented Diffusion Model for Crystal Material Generation}
\author {
    Astrid Klipfel\textsuperscript{\rm 123},
    Yaël Fregier\textsuperscript{\rm 2},
    Adlane Sayede\textsuperscript{\rm 3},
    Zied Bouraoui\textsuperscript{\rm 1}
}
\affiliations {
    \textsuperscript{\rm 1}Univ. Artois, UMR 8188, Centre de Recherche en Informatique de Lens (CRIL), F-62300 Lens, France.\\
    \textsuperscript{\rm 2}Univ. Artois, UR 2462, Laboratoire de Mathématiques de Lens (LML), F-62300 Lens, France.\\
    \textsuperscript{\rm 3}Univ. Artois, UMR 8181, Unité de Catalyse et de Chimie du Solide (UCCS), F-62300 Lens, France.\\
    \{astrid.klipfel, yael.fregier, adlane.sayede, zied.bouraoui\}@univ-artois.fr
}

\usepackage{svg}

\usepackage{amsmath,amssymb,amsfonts}

\def\R{\mathbb{R}}
\def\Z{\mathbb{Z}}
\def\N{\mathbb{N}}
\def\SO{\text{SO}_3(\R)}

\def\SLZ{\text{SL}_3(\Z)}
\def\GL{\text{GL}_3(\R)}

\def\so{\mathfrak{so}_3(\R)}
\def\sl{\mathfrak{sl}_3(\R)}
\def\gl{\mathfrak{gl}_3(\R)}

\def\DKL{D_\text{KL}}
\def\M{\mathcal{M}}
\def\T{\mathbb{T}}
\def\E{\mathbb{E}}
\def\Norm{\mathcal{N}}
\def\intv#1[#2..#3]{\mathopen{#1[}#2\mathrel{{.}\,{.}}\nobreak#3\mathclose{#1]}}
\DeclareMathOperator*{\argmin}{arg\,min}
\newcommand{\module}[1]{{[ #1]}}
\DeclareMathOperator{\Tr}{Tr}

\newtheorem{lemma}{Lemma}

\usepackage{bibentry}

\begin{document}

\maketitle

\begin{abstract}
Discovering crystal structures with specific chemical properties has become an increasingly important focus in material science. However, current models are limited in their ability to generate new crystal lattices, as they only consider atomic positions or chemical composition. To address this issue, we propose a probabilistic diffusion model that utilizes a geometrically equivariant GNN to consider atomic positions and crystal lattices jointly. To evaluate the effectiveness of our model, we introduce a new generation metric inspired by Frechet Inception Distance, but based on GNN energy prediction rather than InceptionV3 used in computer vision. In addition to commonly used metrics like validity, which assesses the plausibility of a structure, this new metric offers a more comprehensive evaluation of our model's capabilities. Our experiments on existing benchmarks show the significance of our diffusion model. We also show that our method can effectively learn meaningful representations.
\end{abstract}

\section{Introduction}
Crystal materials play a crucial role in numerous technological applications, such as the aerospace industry and semiconductors. As global warming poses a significant threat to our society, developing new materials can be vital to our ability to adapt to change and minimize our carbon footprint. To support the ecological transition, it is important to search for and generate new crystals, such as new semiconductors, that can be used to produce solar hydrogen or store hydrogen in solid form \cite{doi:10.1021/acsenergylett.9b02582,RUSMAN201612108}. New methods, such as high-throughput screening, rely heavily on machine learning models to speed up the development process. Recently, there has been growing interest in developing machine learning models to create new crystals  \cite{NOH20191370,Dan2020,Long2021}. In particular, Graph-based Neural Networks (GNNs) models have been proven effective for handling crystal structures. While, however, these models have shown excellent performances for regression tasks and property prediction such as the formation energy or the band-gap \cite{XieandGrossman2018,jorgensen2018neural,Choudhary2021}, the progress is still hampered by the limited extent to which generation is performed.  As far as we know, no existing model can perform sampling with a precise composition. This type of sampling is crucial for exploring the various phases of a material within a specific composition and generating a convex hull. The ability to do this is essential in searching for new materials  \cite{doi:10.1021/acscentsci.0c00426}. Our work is unique in enabling this type of sampling, whereas other models are not as well-suited for this task. In this paper, we propose an effective probabilistic sampling diffusion process with GNN for crystal materials generation.

Crystals are periodic structures consisting of a minimal set of atoms (unit cell) infinitely repeated in all directions of 3D space. Contrary to organic chemistry, where diffusion models have led to spectacular results on generation, a key challenge when dealing with crystals is how the periodicity and the complex chemistry of materials are taken into account and handled by the generative model. In \cite{xie2021crystal}, a GNN-based model for material generation has been proposed where the diffusion process is only applied to atomic positions but not cells. Generating an appropriate cell is crucial for crystals. A cell defines the repetition pattern of a material and its density, which have an important impact on the properties of the crystal lattice. To this end, we propose a diffusion process on all the geometry of the structures (atomic positions and cells). In particular, our model relies on a diffusion process in a torus (for the atoms) and on a lattice (for the cells). 
We use a neural network based on an equivariant vector field to create our lattices, which sets us apart from others who only predict lattice parameters from geometry. Additionally, we use an equivariant graph neural network layer. This layer is equivariant to both $\E$ and $\SLZ$. Although introducing this diffusion process is challenging, we offer an elegant solution and explain how to reverse the process. Our model generates a more realistic lattice, as confirmed by our experiments.

To assess the quality of the generated structure, several metrics such as validity, density and formation energy are commonly used \cite{xie2021crystal,doi:10.1021/acs.jcim.0c00464}. The validity metric performs well when the generated crystals are bad, i.e., to discard invalid materials having a poor validity ratio. When a generative model produces a good crystal structure, this metric is no longer meaningful. The density and formation energy are good metrics. However, they are arbitrary and partially cover material characteristics. 
There exist also metrics that compare the distributions of other physical quantities.
Contrary to the validity metric, the comparison of statistical distributions is a good metric. However, even though the compared quantities are carefully chosen such as density and energy, they are not enough to assess the quality of generated materials. Namely, other quantities such as band-gap, magnetism, piezoelectricity and elasticity are also relevant for characterizing materials.  
To this end, in addition to existing metrics, we add Frechet ALIGNN Distance (FAD), which provides a more complete view for evaluating crystals as it takes into account a large number of features. This metric aims to provide a robust evaluation that is similar to the FID used for computer vision but using ALIGNN \cite{Choudhary2021} instead of inception v3.

The main contributions of this paper are as follows: (i) we propose a diffusion process on the whole geometry of crystal materials; (ii) we introduce a regression model that learns the reverse process and generates materials; (iii) we propose a new metric to evaluate the performance of material generative models; (iv) we performer deep experimental analysis of our models on existing benchmarks; (v) we provide the code to train our proposed model with diffusion and sampling of crystalline structures ; (vi) we provide a tool to calculate the generation metrics as an independent python package \footnote{Data and source code are available at: \url{https://github.com/aklipf/gemsdiff}}.

\section{Related Works}

Recent years have witnessed an important interest in developing machine learning models for material science in both organic chemistry \cite{satorras2021en,igashov2022equivariant,schneuing2022structurebased,pmlr-v139-luo21a,Shi*2020GraphAF,Wu2021.06.06.447297,xu2022geodiff,trippe2023diffusion} and materials science \cite{XieandGrossman2018,jorgensen2018neural,Choudhary2021,klipfel2023equivariant}. Most state-of-the-art generative methods can be divided according to the material representation: fingerprint \cite{doi:10.1021/acscentsci.0c00426,REN2022314,nouira2019crystalgan,Dan2020,https://doi.org/10.1002/advs.202100566,D0CP03508D,sawada2019study,D0CE01714K}, voxel \cite{Long2021,NOH20191370,doi:10.1126/sciadv.aax9324, doi:10.1021/acs.jcim.0c00464,DBLP:journals/corr/abs-1909-00949} and graph-based representation \cite{xie2021crystal,PhysRevMaterials.6.033801,Gibson2022,https://doi.org/10.48550/arxiv.2012.02920}. The fingerprint or voxel-based models are, in general, built upon VAE \cite{REN2022314,Long2021,NOH20191370} and GAN \cite{doi:10.1021/acscentsci.0c00426,Long2021,https://doi.org/10.1002/advs.202100566}. Although these models perform well, generated structures often lack stability and are subject to different constraints. For example, the generated crystals are often limited to cubic cells  \cite{https://doi.org/10.1002/advs.202100566,Long2021,doi:10.1021/acsomega.9b00378,doi:10.1021/acs.jcim.0c00464}, or have specific chemical composition \cite{doi:10.1021/acscentsci.0c00426,Long2021,NOH20191370}. 
With such constraints, a lot of data is required on specific chemical composition and learned knowledge cannot be easily transferred and reused. Moreover, Voxel-based models are, in general, very expensive to train and difficult to use on large volumes of data. Finally, with a diffusion process, we can often have very good performance in generation while avoiding the risk of collapse mode (the main problem with GAN-based models).

Recently, graph-based models with diffusion process have shown impressive results in generation for organic chemistry \cite{schneuing2022structurebased,igashov2022equivariant,Wu2021.06.06.447297}. However, there are few works for crystal generation \cite{xie2021crystal}. 
There are only a few models that can accurately predict lattice while remaining invariant to the $\E$ group \cite{xie2021crystal,yan2022periodic}. However, these models tend to not be invariant to the $\SLZ$ group, which may limit their prediction capabilities.
In \cite{xie2021crystal}, the diffusion process does not apply to cells and only concerns atomic positions and atomic numbers. 
It is not designed to sample a given chemical composition, i.e., sampling structures for various atomic ratios. This is a clear limitation when constructing the convex hull of a given chemical composition. Moreover, the training phase suffers from instability since the diffusion is not clearly defined in the torus or on the cell, which consequently lowers the precision of the predicted lattice parameters. These limitations prevent the application of such diffusion models on real-world problems, i.e., when studying chemical systems for a specific application. Our goal is to create a model that can generate structures with specific chemical compositions but with varying proportions. To accomplish this, our model allows us to perform sampling by adjusting the rate given as input.


\section{Background}
We first introduce background elements about crystalline materials and the probabilistic diffusion process. 

\subsection{Crystalline Materials}
Crystals are structures made up of a minimum number of atoms called unit cell that are repeated infinitely in all directions of space to form a lattice. This lattice can be seen as an atomic point cloud, where an atom may be repeated in multiple positions due to the space tilling. Hence, its local environment can overlap with adjacent repetitions. To define a crystal, we use the atomic positions $x_i$, which are within the range of $[0,1[^3$, the chemical feature space $F$ that represents the chemical information of each atom $z_i\in F$, and the lattice $\rho\in\text{GL}_3(\mathbb{R})$ that represents the periodicity. The infinite point cloud can be expressed as follows:
\begin{equation}
    \langle M\rangle=\{(\rho(x_i+\tau),z_i)|\tau\in\Z^d,i\in\intv[1..n]\}\subseteq\R^d\times F
\end{equation}

Where $\tau$ acts as a $\mathbb{Z}^3$ vector that translates the point cloud. The representation space of all possible materials with $n$ atoms is then defined as $\M^F=\bigcup_{n\in\N}\M^F_n$ with

\begin{equation}
     \M^F_n=\{(\rho,x,z)|\rho\in\GL,x\in[0,1[^{n\times d},z\in F^n\}
\end{equation}

The cloud $\langle M\rangle$ can be rotated and translated by the Euclidian group. However, we can also find an alternative tiling of the space. Let $\langle M\rangle$ invariant in space, which corresponds to an action of $\SLZ$ on the materials, i.e., invariant under the action of the lattice $L = \rho \cdot \mathbb{Z}^d \subseteq \mathbb{R}^d$. 
We recall action groups that a GNN model should satisfy when manipulating crystal lattice \cite{klipfel2023equivariant}: (i) $O(d)$ orthogonal group acting by $g \cdot (\rho, x, z) = (g \cdot \rho, x, z)$; (ii) the translation group $\R^d$ acting by $v \cdot (\rho, x, z) = (\rho, \module{ x + v}, z)$; (iii) the euclidean group $E(d) = \R^d\rtimes O(d)$; (iv) the special linear group $\text{SL}_d(\Z)$  acting by $g' \cdot (\rho, x, z) = (\rho g'^{-1}, \module{ g' x }, z)$. We refer to \cite{klipfel2023equivariant} for more details.

\subsection{Probabilistic Diffusion}

Diffusion models are generative models developed according to the nonequilibrium thermodynamics \cite{pmlr-v37-sohl-dickstein15,NEURIPS2020_4c5bcfec}, which define Markovian diffusion process that incrementally introduces noises to input data, and then learns from a reverse process to distinguish data from noises. While it is relatively easy to add noises to input data, such an operation generally engenders information loss since several identical antecedents to the function that added noises may hold.
The \textit{forward process} or \textit{diffusion process} is defined as a Markov chain that incrementally adds noises to the $x_0$ data. The diffusion process is the \textit{posterior} of a latent variable model and is usually defined using a scheduler of variance $\beta_1,\cdots,\beta_T$:
\begin{equation}
\begin{split}
    q(x_t|x_{t-1}):=&\Norm\big(\sqrt{1-\beta_t}x_{t-1},\beta_t\text{I}\big),\\
    q(x_{1:T}|x_0)=&\prod_{t=1}^{T}q(x_t|x_{t-1})
\end{split}\label{eq:forward_process}
\end{equation}

The \textit{reverse process} and the \textit{joint distribution} of the latent variable with $p(x_T)=\Norm(0,\text{I})$ can similarity be defined as a Markov chain:
\begin{equation}
    \begin{split}
        p_\theta(x_{t-1}|x_t):=&\Norm\big(\mu_\theta(x_t,t),\Sigma_\theta(x_t,t)\big),\\
        p_\theta(x_{0:T})=&p(x_T)\prod_{t=1}^{T}p_\theta(x_{t-1}|x_t)
    \end{split}\label{eq:reverse_process}
\end{equation}
Similar to variational models, diffusion models can be trained by minimizing the evidence lower bound (ELBO): 
\begin{equation}
    \begin{split}
        &\mathcal{L}=\E_q[\underbrace{\DKL\big(q(x_T|x_0)||p_\theta(x_T)\big)}_{L_T}\\
        &+\sum_{t>1}\underbrace{\DKL\big(q(x_{t-1}|x_t,x_0)||p_\theta(x_{t-1}|x_t)\big)}_{L_{t-1}}\underbrace{-\log p_\theta(x_0|x_1)}_{L_0}]
    \end{split}\label{eq:elbo}
\end{equation}

Equation \ref{eq:elbo} can be simplified when the Markov chain is defined with Gaussian distributions and the variance scheduler only depends on hyperparameters.
\begin{equation}
\mathcal{L}_\text{simple}:=\E_q\big[||\tilde{\mu}_t(x_t,x_0)-\mu_\theta(x_t,t)||^2\big]\label{eq:simplifed_diffusion}
\end{equation}

with $\tilde{\mu}_t(x_t,x_0):=\frac{\sqrt{\bar{\alpha}_{t-1}}\beta_t}{1-\bar{\alpha}}x_0+\frac{\sqrt{\alpha_t}(1-\bar{\alpha}_{t-1})}{1-\bar{\alpha}_t}x_t$, $\alpha_t:=1-\beta_t$, $\bar{\alpha}_t:=\prod_{s=1}^{t}\alpha_s$ and $x_t(x_0,\epsilon)=\sqrt{\bar{\alpha}_t}x_0+\sqrt{1-\bar{\alpha}_t}\epsilon$ with $\epsilon\sim\Norm(0,\text{I})$.

\section{Diffusion on a Torus with an Equivariant Graph Neural Network}
\label{seq:method}
We need two steps to apply diffusion to crystal generation. First, we define a diffusion on torus to handle the periodicity of a material. Second, we introduce a model $f_\theta$ that can learn the reverse process. We use an Equivariant Graph Neural Network $f_\theta:\M^F_n\times\N \rightarrow \M^F_n $ that takes as input a crystal $M_t=(\rho,x,z) \in \M^F_n$ at a diffusion step $t\in\N$ and predicts a new cell $\rho’$ as well as new atomic positions $x’$. This approximation of $M_{t-1}$ enables to denoise the step $t$ of the diffusion process. Since a crystal lattice forms a toroidal space, we adapt diffusion to such a space using a loss of two parts: A first part concerns the shape of the torus $\mathcal{L}^\rho$, and the second part concerns the diffusion inside the torus $\mathcal{L}^x$. The final loss is the sum of the two parts: 
$\mathcal{L}:=\mathcal{L}^\rho+\mathcal{L}^x$. 
We now describe these different steps in more detail.

\subsection{Equivariant Graph Neural Network}
\label{seq:equivariant_gnn}
We start by introducing the map $f_\theta$ which is an equivariant GNN that takes a material and a diffusion step as input and outputs a new material with the same chemical composition but an updated lattice $\rho$ and atomic position $x$. 
\begin{align*}
f_\theta:\M^\mathbb{A}_n\times\N & \to \M^\mathbb{A}_n\\
    (\rho,x,z,t)                          & \mapsto(\rho',x',z)
\end{align*}

The goal of the map $f_\theta$ is to enable inverting the diffusion process one step at a time by parameterizing the $p_\theta(M_{t-1}|M_t)$. This map is then defined with a step-dependent auto-encoder GNN.
\begin{align*}
    \text{GNN}_{AE}: \GL\times\T^n\times F^n\times \N & \to\GL\times\R^{n\times3}\times F^n \\
    (\rho,x,z,t)                          & \mapsto(y^\rho,y^x,y^z)
\end{align*}

The pair of action $(y^\rho,y^x)$ predicted by the GNN can be used to update the geometry of the material as follows:
\begin{equation}
    (\rho',x',z)=(y^\rho,y^x,\text{id})\cdot(\rho,x,z)\text{ with }\begin{cases}
        \rho':=y^\rho\rho \\
        x'_i:=[x_i+y^x_i].
    \end{cases}
\end{equation}

The auto-encoder defining the denoising task consists of the composition of three maps: $$\text{GNN}_{AE}:=\text{Decod}\circ \text{Encod}\circ \text{emb}.$$
The first map, $\text{emb}$, embeds the atoms for the network. This embedding is built from the atomic number and the diffusion step as 
$\text{emb}:\mathbb{A}\times\N\to F$, given by the following two cases:
\begin{equation}
    \begin{cases}
        \text{emb}(z,t)_{2k}   & =\sin(\frac{t}{n^{2k/d_\text{emb}}})+\text{emb}_{z,2k}   \\
        \text{emb}(z,t)_{2k+1} & =\cos(\frac{t}{n^{2k/d_\text{emb}}})+\text{emb}_{z,2k+1}
    \end{cases}
\end{equation}

With $\text{emb}_z$ is a vector associated with the atomic number $z$. The embedding works by associating a vector to each atomic number and adding $t$, the index of the diffusion step, to this representation in the same way that positional encoding works in a transform on tokens. The remaining maps $\text{Encod}$ and $\text{Decod}$ are built with components of $\text{GNN}_\text{enc}$ and $\text{GNN}_\text{dec}$, two equivariant GNN architectures of the type \cite{Klipfelfsb2023}. These components are combined as follows:
\begin{subequations}
    \label{eq:gemsdiff_autoencodeur}
    \begin{equation}
        z^\text{emb}_i=\text{emb}(z_i,t)
    \end{equation}
    \begin{equation}
(\cdot,\cdot,z^\text{enc})=\text{GNN}_\text{enc}(\rho,x,z^\text{emb})
    \end{equation}
    \begin{equation}
     (\rho',x',\cdot)=\text{GNN}_\text{dec}(\rho,x,z^\text{enc})\cdot(\rho,x,\cdot)
    \end{equation}
\end{subequations}
We show that $f_\theta$ is equivariant (proof in Appendix A).
\begin{equation}
    f_\theta(g\cdot M,t)=g\cdot f_\theta(M,t) | \forall g\in\text{E}(3)\times\SLZ.
\end{equation}

In practice, we will use this architecture, but we decode $\rho'$ from a cubic cell of volume 1 rather than from $\rho$ because we found that more stable and produces better results.  Notice that while this modification breaks the equivariance between the input and the output of the network, the GNN that constitutes it remains equivariant. As such, we still benefit from the GNN structuring factor. So, $\rho'$ is computed as follows:
\begin{equation}
    (\rho',x',\cdot)=\text{GNN}_\text{dec}(\rho,x,z^\text{enc})\cdot(\text{I}_3,x,\cdot).
\end{equation}

\subsection{Diffusion Process for Crystals}
\label{seq:lattice_diffusion}
We first propose a diffusion on $\GL$, the space of lattices. As the space is non-linear, we linearize with logarithm to use the diffusion process of $\R^n$ and bring back the result of the diffusion to $\GL$ using the exponential map. We then show that the usual diffusion process in the Euclidian space induces a diffusion process on the torus, enabling us to also apply diffusion to atoms within a lattice.

\noindent\textbf{Lattice Vector Space.} One can see a cell $\rho\in\GL$ as the image by the exponential map of an element $g\in\gl$, i.e. as $\rho=\exp(g)$. By choosing a supplement $\sl\setminus$ of $\so\subset\sl$, one can decompose the Lie algebra $\gl$ as a direct sum:
\begin{equation}
\gl=\so\oplus\sl\setminus\so\oplus\mathfrak{1}_3,
\end{equation}
with
\begin{itemize}
    \item $\so\subset\gl$: rotation algebra. 
    \item $\sl\setminus\so\subset\gl$:  a subspace of the special linear algebra supplementary to $\so$. One can choose its elements to be responsible for the shape of the cell but with a cell of volume 1 (det 1).
    \item $\mathfrak{1}_3\subset\gl$: the algebra composed of the identity matrix singleton, which generates the volume of the cell.
\end{itemize}

This decomposition of the linear Lie algebra allows us to define a lattice vector space where vectors are in 9 dimensions with 3 dimensions controlling the orientation of the cells, 5 dimensions for the shape of the cells and 1 dimension for volume. Since linear spaces are of the same finite dimension, they are isomorphic. A linear isomorphism is given by choosing a basis in $\gl$. An explicit basis $ \pi\in\gl^9$ defining the following isomorphism is given in Section B of the Appendix. Notice that this process is invertible. 
\begin{equation}
    \begin{aligned}
        \pi: \R^9 & \to\gl       \\
        x         & \mapsto\pi x.
    \end{aligned}\text{\hspace*{0.5cm}}\label{eq:mapping_vetor_space}
\end{equation}

\noindent\textbf{Lattice Diffusion.} Based on Equation \ref{eq:simplifed_diffusion} and \ref{eq:mapping_vetor_space}, we can define the loss in the lattice vector space where $x^\rho_0$ is the vector of original lattice $\rho_0$ and $x_\theta(M_t,t)$ is the predicted lattice vector, resulting in: 
\begin{equation}
\begin{split}
\mathcal{L}:=&\E_q\big[||\tilde{\mu}_t(x^\rho_t,x^\rho_0)-\mu_\theta(M_t,t)||^2\big]\\
=&\E_q\big[\frac{\sqrt{\bar{\alpha}_{t-1}}\beta_t}{1-\bar{\alpha}}||x^\rho_0-x_\theta(M_t,t)||^2\big]
\end{split}\label{eq:loss_lattice_algebra}
\end{equation}

Finally, we cannot directly use the loss defined in Equation \ref{eq:loss_lattice_algebra} as the operations $\exp$ and $\log$ do not have a well-defined gradient in Pytorch due to the Eigenvalues decomposition. We, therefore, define a loss of lattice reconstruction that is invariant to the actions of $\SO$, which has the same minimum as the Equation \ref{eq:loss_lattice_algebra} from the lattice predicted by the model $\rho_\theta(M_t,t)$.
\begin{equation}
\mathcal{L}^\rho:=\E_q\big[d(\exp(\pi\tilde{\mu}_t(x^\rho_t,x^\rho_0)),\rho_\theta(M_t,t))]
\end{equation}

We choose as distance measure between the two lattices $d(\rho,\rho')=||p(\rho)-p(\rho')||$ with $p:\GL\to\R^6$ as an application that associates the normalized crystallographic lattice parameters $(a,b,c,\alpha,\beta,\gamma)$ to a lattice $\rho$.

\noindent\textbf{Torus Diffusion.} We now detail the forward process inside a torus and the loss function. The principle is similar to the classical diffusion process but with a periodic space.

\textit{Forward Process Inside a Torus:} We define the torus as a quotient space such that $\T=\R^d/\Z^d$ where we consider the embedding of $\Z^d$ in $\R^d$ given by integer coordinates. For a distribution $P$ on $\R^d$ of density $p$, we use $[P]_\T$ to denote the distribution $P$ in the torus with density $p_\T(x)=\sum_{\tau\in\Z^d}p(x+\tau)$. We show that if $P$ is a density, then $[P]_\T$ is a density (proof in Appendix C). 

\begin{lemma}
$[\Norm(\mu,\sigma\text{I})]_\T\xrightarrow[\sigma\to+\infty]{}\mathcal{U}(0,1)^d$
\label{lemma:limit_diffusion}
\end{lemma}

The proof of Lemma \ref{lemma:limit_diffusion} is provided in Appendix D. Consequently, we choose $p_\theta(x_T)=\mathcal{U}(0,1)$. We can now introduce our forward process with $q$ such that.
\begin{equation*}
     q(x_t|x_{t-1}):=[\Norm(x_0,\beta_t)]_\T \text{ then }
\end{equation*}
\begin{equation}
    q(x_t|x_0)=[\Norm(x_0,(1-\bar{\alpha}_t)\text{I})]_\T
\end{equation}

Notice that there is no need to shift the norm because the distribution tends towards a uniform distribution. Now the reverse process with $p_\theta$ in a similar way to equation \ref{eq:reverse_process}:
\begin{equation}
    p_\theta(x_{t-1}|x_t):=[\Norm(\mu_\theta(x_t,t),\sigma^2_t\text{I})]_\T
\end{equation}

We deduce that we can sample $x_{t-1}$ from $x_t$ and $x_0$ directly as $q(x_{t-1}|x_t,x_0)=[\Norm(x_0,\tilde{\beta}_t\text{I})]_\T$

\textit{Loss Function:} The loss is composed of 3 parts from equation \ref{eq:elbo}: $L_T$, $L_{t-1}$ and $L_0$. $L_T$ is a constant because it depends only on $\beta_t$ (variance scheduler). $L_0$ is the same as $L_{t-1}$ in our context. To perform the last diffusion step, we use $L_{t-1}$ with a variance of $0$. This leaves us $L_{t-1}$:
\begin{equation}
    L_{t-1}=\E_q\big[\DKL(q(x_{t-1}|x_t,x_0)||p_\theta(x_{t-1}|x_t))\big]\label{eq:L_tminue1}
\end{equation}

Being periodic, the distributions $q$ and $p_\theta$ can shift by $k\in\Z^3$. We then obtain Equation \ref{eq:L_tminue1} which is equivalent to the Equation \ref{eq:L_tminue1_torus}. Proofs are given in Appendix E.
\begin{equation}
    L^k_{t-1}:= \E_{q,k}\big[\frac{1}{2\sigma^2_t}||x_0-x_t-k-\epsilon_\theta(x_t,t)||^2\big]\label{eq:L_tminue1_torus}
\end{equation}


We can now define the final loss. As optimizing for any $k$ minimizes \ref{eq:L_tminue1}, we choose the $k$ that minimizes the trajectory performed by the atom. We then attend the targeted position following the shortest path. 
 
 
\begin{equation}
    \mathcal{L}^x:=\E_q\big[||x_0-x_t-k^*-\epsilon_\theta(x_t,t)||^2]
\end{equation}
\begin{equation*}
    \text{ with } k^*=\argmin_{k\in\Z^d}||x_0-x_t-k||
\end{equation*}

Finally, we obtain a reconstruction loss where the network is trained to move atoms back to their original position following the shortest possible path.


\begin{table}
\centering
\footnotesize
    \begin{tabular}{l|c}
    \toprule 
        Parameter & Value\\
        \midrule
        batch size & 128 \\
        epochs & 512 \\
        learning rate & 1e-3\\
        knn & 32\\
        diffusion steps & 100\\
        $(\beta^\rho_0, \beta^\rho_T)$ & (1e-5, 1e-1)\\
        $(\beta^x_0, \beta^x_T)$ & (1e-6, 2e-3)\\
        \bottomrule
    \end{tabular}
    \caption{Hyperparameters of GemsDiff}
    \label{tab:hyperparameters}
\end{table}

\section{Experiments}
We present an evaluation of our proposed models GemsDiff
We will, in particular, focus on experiments about diffusion and evaluation based on Frechet distance. All experiments were conducted on a subset of Materials Project \cite{jain2013commentary}, which is a dataset that contains crystalline materials that have been studied with ab-initio calculations (physical simulation). We follow the same setting as  \cite{xie2021crystal} and filter the dataset by only keeping stable materials having less than 20 atoms.
We only keep stable materials as we aim to assess the capability of our model to produce stable structures. 
We use the same training, validation and test splits as \cite{xie2021crystal} for comparable results.

Notice that in our research, we have chosen not to evaluate Perov-5 and Carbon-24 for various reasons. Perov-5 is a dataset of perovskite structures with identical atomic positions but different compositions, which does not align with the purpose of our research, which is to predict geometry based on composition. Carbon-24 is a structured dataset simulated at 10GPa, which is equivalent to 100k atmospheric pressure, and prediction models from the literature cannot accurately predict the properties of materials under such different conditions. Lastly, our focus is on semiconductors, and we are not interested in carbon structures.

\noindent\textbf{Training} The different hyperparameters we used for training are reported in Table \ref{tab:hyperparameters}. The training is performed following Algorithm \ref{alg:training}, which uses the GNN and loss functions provided in Section \ref{seq:method}. The training time is estimated to 36 hours on an Nvidia RTX 8000 GPU with 48Go. The sampling of the test set, which contains 9046 crystals, takes about 45 minutes, which is normal for a diffusion model.

\noindent\textbf{Baselines.} We evaluate GemsDiff against the state-of-the-art baselines: CDVAE \cite{xie2021crystal}, FTCP \cite{REN2022314}, PGCGM \cite{Zhao2023}, G-SchNet\cite{NEURIPS2019_a4d8e2a7} and P-G-SchNet \cite{NEURIPS2019_a4d8e2a7}. We re-trained the CDVAE and FTCP to sample new crystals and calculate the Frechet ALIGNN distance (FAD) (as described in section 5.2). For PGCGM \cite{Zhao2023}, we used a pre-trained model that allows the generation of crystal materials. For G-SchNet and P-G-SchNet, we report results from \cite{xie2021crystal}.

\noindent\textbf{Sampling process.} To generate structures with GemsDiff, we use chemical compositions of the test set. As such, geometrical information and chemical compositions have not been seen by the models. We use the Langevin dynamics with Algorithm \ref{alg:sampling} following the same parameters as in the training. We generate a dataset having the same size as the test set to be able to compare them. The data generation is performed from the checkpoint, which has the best loss in validation. We use an exponential moving average of the weights with a decay of $0.995$. Then, we use a set of crystals generated without filtering and without optimizing the geometry (without simulation methods to improve the geometry). We also do not search the space group or symmetries of the structures to improve geometry. 

\begin{algorithm}[t]
\footnotesize
\caption{Training}\label{alg:training}
\textbf{Input:} $(\rho_0,x_0,z)\in\M$
\begin{algorithmic}[1]
\REPEAT
\STATE $t\sim\mathcal{U}(\{1,\cdots,T\})$
\STATE $x^\rho_t\sim q(x^\rho_t|\pi^{-1}\log\rho_0)$
\STATE $x_t\sim q(x_t|x_0)$
\STATE $(\rho_\theta, \epsilon_\theta)\gets f_\theta(\exp\pi x^\rho_t, x_t, z, t)$
\STATE gradient descent step on $\nabla\mathcal{L}$
\UNTIL{converged}
\end{algorithmic}
\end{algorithm}
\begin{algorithm}[t]
\footnotesize
\caption{Sampling}\label{alg:sampling}
\textbf{Input:} $z\in\mathbb{A}^n$
\begin{algorithmic}[1]
\STATE $x_T \sim \mathcal{U}(0,1)$
\STATE $\rho_T\gets \exp(\pi x^\rho_T) \text{ ,} x^\rho_T\sim \Norm(0,1)$
\FOR{$t\gets T,\cdots,1$}
    \STATE $(x', \rho')\gets f_\theta(\rho_t, x_t, z, t)$
    \STATE $x_{t-1}\gets[x'+\sigma^x_t\epsilon] ,\epsilon \sim \Norm(0,1)$
    \STATE $\mu_\rho\gets \tilde{\mu}_t(\pi^{-1}\log\rho_t,\pi^{-1}\log\rho')$
    \STATE $\rho_{t-1}\gets\exp\big(\pi(\mu_\rho+\sigma^\rho_t\epsilon)\big) ,\epsilon \sim \Norm(0,1)$
\ENDFOR
\end{algorithmic}
\end{algorithm}

\begin{figure}[t]
    \centering
    \includesvg[width=0.8\columnwidth]{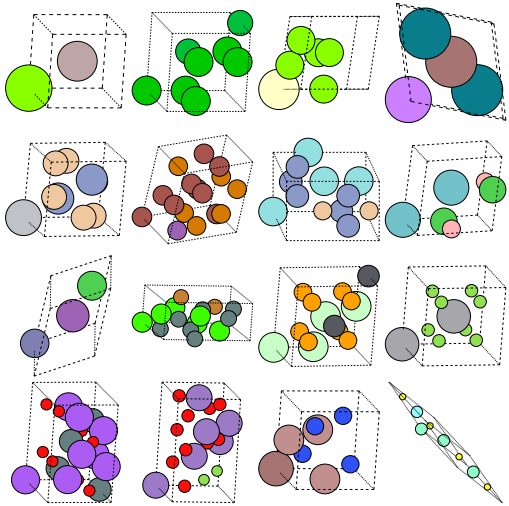}
    \caption{Sampling with GemsDiff and Langevin Dynamics}
    \label{fig:diffusion}
\end{figure}

\begin{figure}[t]
    \centering
    \includesvg[width=0.8\columnwidth]{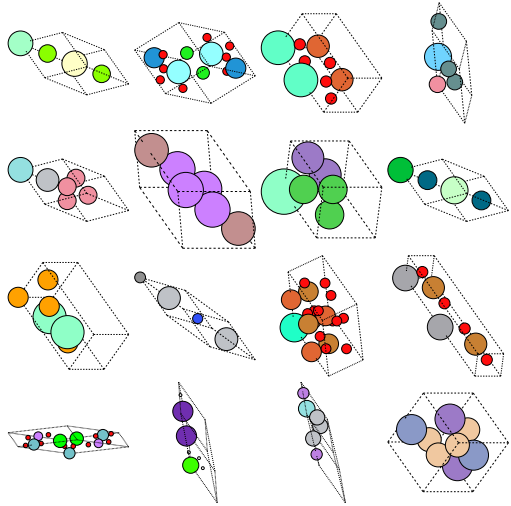}
    \caption{Crystals from materials project}
    \label{fig:mp-20}
\end{figure}

We can see structures generated by GemsDiff and from Materials Project in Figures \ref{fig:diffusion} and \ref{fig:mp-20}. We can observe several well-known lattice systems having different constraints:  cubic cell, monoclinic, orthorhombic, tetragonal, and, more importantly, hexagonal with 60° and 120° angles as lattice parameters. It is important to notice that there is no explicit rule in our model that facilitates the generation of such constraints in the lattice parameters. Interestingly enough, many lattices fall in a known system with a precision of less than 0.5 Angstrom and 0.1 Angstrom. Moreover, we can see that atoms of the materials make patterns like alignment, cubic geometry, or circular ring composed of 6 atoms. One can observe that the geometrical shape are coherent, i.e. often composed of the same type of elements. 
We stress the fact that there are no explicit constraints encoded in our model to encourage obtaining such structures. It is an emerging behavior that appears during training. Moreover, the observed characteristic of real materials is present in the generated data without any explicit bias encoded in our model. This offers an excellent indicator of the quality of the generated structures. 


\begin{table*}
    \centering
    \footnotesize
    \begin{tabular}{l|ccc}
    \toprule
        Method & Validity (\%) & density (EMD $g/cm^3$) & energy (EMD $eV/atom$) \\
        \midrule
        FTCP$^*$ & 88.18 & 1.586 & 0.8281\\
        PGCGM & 99.03 & 0.89 & 1.51\\
        G-SchNet & 99.65 & 3.034 & 42.09\\
        P-G-SchNet & 77.51 & 4.04 & 2.448\\
        CDVAE & \textbf{100.0} & 0.6875 & 0.2778\\
        GemsDiff & 99.05$\pm$0.20 & \textbf{0.3331$\pm$0.0327} & \textbf{0.1984$\pm$0.0086  } \\
        \bottomrule
    \end{tabular}
    \caption{Results for different models in terms of validity, density and energy. For FTCP baseline, we report the results that we obtained, which are better than the original value reported in \cite{xie2021crystal}. }
    \label{tab:metrics_previous_works}
\end{table*}

\subsection{Evaluation on Validity, Earth Mover distance (EMD) for Density and Energy}
To assess the quality of the generated crystals, we first rely on existing metrics: Validity and Earth Mover distance (EMD) for density and energy. EMD is calculated from the structure generated without filtering where the same amount of structures as in the test set (9048) is generated. 
Validity \cite{doi:10.1021/acs.jcim.0c00464} refers to the percentage of structures that don't have two atoms with a relative distance less than 0.5 Angstrom. This metric evaluates the percentage of structures that are incoherent as physically having a distance lower than 0.5 Angstrom is impossible. Even if it is not a good metric, it at least allows us to spot the badly generated crystals. EMD, also called Wasserstein distance (or Wasserstein 1), allows the comparison of the distance between two statistical distributions. A generative model should be able to generate a structure distribution with properties that are close to its training distribution. So, the farther the distance between the distribution of properties of the generated structures to the distribution of the trained properties, the more different the generated data from the trained data. Consequently, the smaller the EMD, the better the result. We consider density distribution and energy distribution for the EMD metrics. As it is difficult to obtain and very expensive to simulate, some GNNs allow us to estimate it. We can use the same GNN for the Fréchet distance (ALIGNN) to estimate the energy (Section 5.2). 
The results are reported in Table \ref{tab:metrics_previous_works}. Our model outperforms the baselines on all the properties. EMD of the formation energy is significantly better than CDVAE. We also notice that the biggest gain is made for the density of the generated structures.  We can explain this by the fact that our model does diffusion on the cell as well as on the atomic positions. In this case, our model  generates a volume that is more adapted to the composition of the structures. The validity is lower for our model than for CDVAE, but this metric is limited and our model remains very competitive.

\subsection{Frechet Distance with ALIGNN (FAD)}

\begin{figure}[t]
    \centering
    \includesvg[width=0.9\columnwidth]{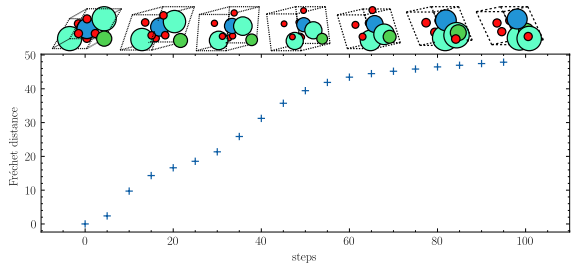}
    \caption{FAD and diffusion step. 
    }
    \label{fig:fad_diffusion}
\end{figure}

We generate a vector of the latent space with a GNN (ALIGNN) to which we remove the last prediction layer and we compute the Frechet distance as in \cite{NIPS2017_8a1d6947}. We use a pre-trained model. Table \ref{tab:experiments_fad} contains all FAD for the generated data. 
To test the relevance of the Frechet distance, previous works on computer vision test the generated data with distortion or with different levels of scattering. In the case of materials, distortion and scattering are very similar. We perform a test for diffusion because it will also allow us to make a link between the FAD and the distance of the generated structures with equilibrium positions. To this end, we apply several levels of diffusion to see the associated FAD. We recall that the lower the FAD, the more similar the distributions are. So the lower the FAD, the more the set of tested materials contains materials close to their equilibrium. As we can see in Figure \ref{fig:fad_diffusion}, the FAD is almost zero for $t=0$, i.e. without diffusion. In contrast, the more $t$ increases, the more the FAD increases. From a certain level of scattering, the FAD reaches a plateau and does not progress anymore. This result is interesting because we observe a behavior close to the FID in the diffusion models used in computer vision \cite{NEURIPS2020_4c5bcfec}. We also notice that the FAD reaches a lower plateau, i.e. the values are globally lower than for the FID in computer vision. This difference can be simply explained by the fact that the vectors used in inceptions are of size 2048 while the latent vectors produced by ALIGNN are only of size 256.
\begin{table}
\footnotesize
\centering
    \begin{tabular}{l|c}
    \toprule
        Method & Fréchet Distance \\
        \midrule
        PGCGM & 30.692\\
        FTCP & 16.562\\
        CDVAE & 1.777\\
        GemsDiff & \textbf{1.509$\pm$0.115}\\
        \bottomrule
    \end{tabular}
    \caption{
    To ensure consistent crystal quality in our training process, we repeat the training eight times and calculate the mean and standard deviation. This computation is carried out on 9048 structures without any filtering.}
    \label{tab:experiments_fad}
\end{table}

\begin{figure}[t]
    \centering
    \includesvg[width=0.9\columnwidth]{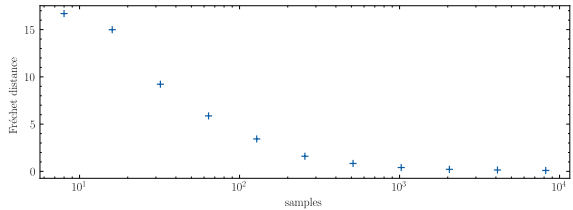}
    \caption{FAD for multiple sample sizes}
    \label{fig:fad_samples}
\end{figure}

To validate the number of instances between two samples, we compute the FAD by progressively increasing the sample size. We evaluate how the metric behaves and try to see when the FAD is close enough to 0. This is a trick that has already been done for the FID, and it allows us to be sure that the metric will not be too disturbed by statistical fluctuations. We can see the results for several sample sizes in figure \ref{fig:fad_samples}. As for the FID, we can see that about 10000 samples are largely sufficient to have a reliable FAD measurement. So we will compare with sample sizes of 9046 (test set).
In Table \ref{tab:experiments_fad}, we can see that our model has the best result for the Frechet distance. This result consolidates the results of the other table with a new metric. To allow a better justification of our architecture, an ablation test is proposed in Appendix F.

\subsection{Discussion} 

Our work has the focus of prioritizing specialized tasks instead of generating compositions. Unlike other models that aim to create stable structures without composition constraints, our model is less suited for this purpose. However, chemists require models that can generate structures with certain compositions as they seek to find different meta-stable phases for a specific composition. Our model is specialized in generating compositions and, therefore, is more suitable for this type of task. Additionally, chemists are interested in determining the convex hulls of certain chemical compositions, which our model is very suitable for. Constraints on chemical composition, such as excluding rare and expensive elements, are also important considerations for chemists. Although some models are good at generating chemical compositions, it can be positive not to be able to generate compositions in certain cases.
By breaking down the lattice in a Lie algebra and producing the lattice from an invariant vector field, the generation of the lattice appears to be enhanced. As Table 3 shows, there is a significant improvement in density, making it much more comparable to the initial distribution. In the future, decomposing the lattice could have other applications. For instance, it may be feasible to construct structures that are part of a particular space group and guarantee that the shape of the lattice is compatible with that space group using characteristics of the vector that represents the lattice in the Lie algebra's space vector.
The FAD aims to create a standardized metric for comparing crystal features, rather than relying on manually selected statistics. The features used are generated by the highly accurate ALIGNN, making them relevant for predicting material properties. However, assessing the consistency of the model's generated features is challenging if ALIGNN is retrained. To address this, we suggest using a pre-trainer version with consistent parameters and features for FAD calculation. To evaluate FAD quality, we can compare it to other metrics for generative model evaluation. Our experimental results show that FAD ranking aligns with other metrics, indicating its relevance for measuring generation quality.

\section{Conclusion}
We have introduced a generative model for crystal materials with a diffusion process adapted to periodic structures based on an equivariant architecture that allows us to learn the reverse process. Our diffusion process performs on the whole crystal geometry (atomic position and cells) allowing us to improve the density prediction and hence obtain more realistic lattices compared to a simple model that predicts lattice parameters.  Our diffusion model is particularly adapted to convex envelope sampling tasks, which is an important issue in materials science \cite{GUBAEV2019148,PhysRevMaterials.2.103804}. Finally, we have also proposed a new metric for evaluating crystal geometry inspired by works on computer vision which are based on a Frechet distance and  diffusion model.  Our model produces interesting generated structures, each of which has a FAD that suggests its closeness to the equilibrium position (relatively stable structure). 

\section*{Acknowledgments}
This work has been supported by ANR-22-CE23-0002 ERIANA, ANR-20-THIA-0004 and by HPC resources from GENCI-IDRIS (Grant 2022-[AD011013338] and 2021-[AD011013288]). 

\bibliography{aaai24}

\appendix
\section*{Appendices}
\addcontentsline{toc}{section}{Appendices}
\renewcommand{\thesubsection}{\Alph{subsection}}

\subsection{Equivariant Graph Neural Network}

We start by recalling that GemsNet\cite{Klipfelfsb2023} acts on a material $(\rho,x,z)$ as follows:

\begin{equation*}
\begin{aligned}
    \text{GNN}: \GL\times\T^n\times F^n & \to\GL\times\R^{n\times3}\times F^n \\
    (\rho,x,z)                          & \mapsto(y^\rho,y^x,y^z)
\end{aligned}
\end{equation*}
\begin{equation*}
    (y^\rho,y^x,y^z)\cdot(\rho,x,z)=(\rho',x',z')\text{ with }\begin{cases}
        \rho':=y^\rho\rho \\
        x'_i:=[x_i+y^x_i] \\
        z'_i:=y^z_i.
    \end{cases}
\end{equation*}

Recall also that GemsNet is based on GemNet\cite{klicpera2022gemnet}, which is an invariant regression model. Consequently, $y^z$ is invariant to all the actions from the Euclidean group and from $\SLZ$ since this architecture relies on invariant properties of the graph under the action of these groups as shown in \cite{klipfel2023equivariant}. Finally, we recall that $y^x$ and $y^\rho$ are defined as:

\begin{equation*}
    y^x_i = \sum_{\gamma\in\mathcal{N}(i)}{w_\gamma}\frac{-e_\tau}{r_\tau}
\end{equation*}
\begin{equation*}
    y^\rho = I_3+\frac{1}{|\Gamma_2|}\sum^{L}_{l=1}\sum_{(\gamma,\gamma')\in\Gamma_2}{w^{l\intercal}_{\gamma\gamma'}}\lambda_{\gamma\gamma'}
\end{equation*}

with $w_\gamma$ and $w^{l\intercal}_{\gamma\gamma'}$ two invariant vectors predicted by GemsNet and $\lambda_{\gamma\gamma'}\in\R^{N_\text{fields}\times 3\times 3}$ a vector field that allows to define how the GNN acts on crystal lattices. This vector field is obtained from geometric invariants by taking their gradients :

\begin{equation*}
\lambda_{\gamma\gamma'}(M) = \frac{\partial r_\gamma}{\partial \rho} \oplus \frac{\partial r_{\gamma'}}{\partial \rho} \oplus \frac{\partial \theta_{\gamma\gamma'}}{\partial \rho}.
\end{equation*}

Now, we summarise how the translation group $\R^3$, the orthogonal group $O(3)$ and re-tiling group $\SLZ$ act on $y^x_i$ and $y^\rho$:

\begin{itemize}
    \item Translation $g\in\R^3$: $g\cdot y^x_i=y^x_i$ and $g\cdot y^\rho=y^\rho$ because $r_\gamma$, $e_\gamma$ and $\rho$ are invariant to translation
    \item Rotation and reflection $g\in O(3)$: $g\cdot y^x_i=y^x_i$ and $g\cdot y^\rho=gy^\rho g^{-1}$ because $r_\gamma$ and $e_\gamma$ are invariant but $g\cdot\lambda_{\gamma\gamma'}=g\lambda_{\gamma\gamma'}g^{-1}$
    \item Re-tiling $g\in\SLZ$: $g\cdot y^x_i=[gy^x_i]$ and $g\cdot y^\rho=y^\rho$ because $r_\gamma$ and $\lambda_{\gamma\gamma'}$ are invariant, but $g\cdot e_{ij\tau}=g\cdot(x_j-x_i+\tau)=[g(x_j-x_i+\tau)]=[ge_{ij\tau}]$
\end{itemize}

We now check that GemsNet is equivariant to the action of $\text{E}(3)\times\SLZ$. Note that as the actions of each subgroup commute with the actions of the other subgroups as shown in \cite{klipfel2023equivariant}, it suffices to prove the equivariance of GemsNet with each individual subgroup to obtain the equivariance with $\text{E}(3)\times\SLZ$.

\begin{itemize}
    \item $g\in\R^3$: $x'(g\cdot M)=[g\cdot x+g\cdot y^x]=[[x+g]+y^x]=[(x+y^x)+g]=g\cdot x'(M)$
    \item $g\in O(3)$: $\rho'(g\cdot M)=(g\cdot y^\rho)(g\cdot\rho)=gy^\rho g^{-1}g\rho=gy^\rho\rho=g\cdot\rho'(M)$
    \item $g\in\SLZ$: $\rho'(g\cdot M)=(g\cdot y^\rho)(g\cdot\rho)=y^\rho\rho g^{-1}=(y^\rho\rho)g^{-1}=g\cdot\rho'(M)$
    \item $g\in\SLZ$: $x'(g\cdot M)=[g\cdot x+g\cdot y^x]=[g(x+y^x)]=[gx'(M)]=g\cdot x'(M)$
\end{itemize}

Consequently, we obtain $x'(g\cdot M)=g\cdot x'(M)\ \forall g\in \text{E}(3)\times\SLZ$ and  $\rho'(g\cdot M)=g\cdot\rho'(M) \ \forall g\in \text{E}(3)\times\SLZ$. So, the defined GNN is equivariant to the $\text{E}(3)\times\SLZ$ action on a material $M$. Regarding the autoencoder, we have:

\begin{equation*}
\begin{cases}
    z^\text{emb}_i&=\text{emb}(z_i,t)\\
    (\cdot,\cdot,z^\text{enc})&=\text{GNN}_\text{enc}(\rho,x,z^\text{emb})\\
    (\rho',x',\cdot)&=\text{GNN}_\text{dec}(\rho,x,z^\text{enc})\cdot(\rho,x,\cdot).
\end{cases}
\end{equation*}

First $z^\text{enc}$ is invariant,  meaning that $z^\text{enc}=(\text{GNN}_\text{enc}\circ\text{emb})(g\cdot M)\ \forall g\in\text{E}(3)\times\SLZ$. As $z^\text{enc}$ is invariant and $\text{GNN}_\text{dec}$ is equivariant, the whole autoencoder is equivariant.

\subsection{Mapping from Lattice to Vector Space}

First, we recall that the mapping between the vector space and the lattice can be defined with $\pi$, the mapping described in Section 4.2, and the following matrix exponential. 

\begin{equation*}
    \rho=\exp(\pi x^\rho).
\end{equation*}
There exist multiple choices of $\pi$ which are equivalent. We consider $\pi$ such that: $\pi_0=\scriptsize\begin{pmatrix}0&1&0\\-1&0&0\\0&0&0\end{pmatrix}$, $\pi_1=\scriptsize\begin{pmatrix}0&0&1\\0&0&0\\-1&0&0\end{pmatrix}$, $\pi_2=\scriptsize\begin{pmatrix}0&0&0\\0&0&1\\0&-1&0\end{pmatrix}$, $\pi_3=\scriptsize\begin{pmatrix}0&1&0\\1&0&0\\0&0&0\end{pmatrix}$, $\pi_4=\scriptsize\begin{pmatrix}0&0&1\\0&0&0\\1&0&0\end{pmatrix}$, $\pi_5=\scriptsize\begin{pmatrix}0&0&0\\0&0&1\\0&1&0\end{pmatrix}$, $\pi_6=\scriptsize\begin{pmatrix}1&0&0\\0&-1&0\\0&0&0\end{pmatrix}$, $\pi_7=\scriptsize\begin{pmatrix}0&0&0\\0&1&0\\0&0&-1\end{pmatrix}$, $\pi_8=\lambda\scriptsize\begin{pmatrix}1&0&0\\0&1&0\\0&0&1\end{pmatrix}$

Such a choice of the basis of the vector space is useful to better control certain features (e.g. orientation, volume, shape) of the lattice during the diffusion process. $\pi_8$ is scaled by $\lambda$ to better control  the volume of the cell during the diffusion process. Indeed, the volume of the lattice is given by $\exp{\Tr{(\pi x^\rho)}}=\exp{\Tr{(\pi_8 x^\rho_8)}}=\exp{(3\lambda x^\rho_8)}$. To prevent the volume from reaching a very low or very high value, $\lambda$ is set to $\frac{0.25}{\sqrt{3}}$. Moreover, the first 3 components of the vectors can be used to make the orientation of the lattice match during the diffusion process.

We can now define a mapping from the lattice space back to the vector space. We use the fact that $\exp$ is locally invertible with the matrix logarithm. It is then possible to find a decomposition of a cell on this basis thanks to the eigenvectors $V$ and the eigenvalues $\lambda$ such that $\rho=V\lambda V^{-1}$ by defining the logarithm of the matrices such that $\log{\rho}=V\log\lambda V^{-1}$. The eigen decomposition can always hold because $\rho\in\GL$. We thus obtain:
\begin{equation*}
 x^\rho\equiv\pi^{-1}\log(\rho)\text{ with } \pi^{-1} \text{ the reverse mapping of } \pi
\end{equation*}

\subsection{Density on a Torus}

We show that if $p$ is a probability density, then $p_\T$ is also a probability density. First we can see that $p_\T\geq0$ because $f_i(x)\geq0\implies\sum f_i(x)\geq0$. Let us calculate the integral of $p_\T$ over the torus $\T$:
\begin{multline*}
    \int_{\T}p_\T(x)dx=\int_{\T}\sum_{\tau\in\Z^d}p(x+\tau)dx\\=\sum_{\tau\in\Z^d}\int_{\T}p(x+\tau)dx=\int_{\R^d}p(x)dx=1.
\end{multline*}
We can conclude that $p_\T$ is a probability density. Also, we can observe that a probability distribution can be arbitrarily shifted by any $\tau\in\Z^d$ and remains the same resulting in lemma \ref{lemma:shift_torus}
\begin{lemma}
    $\forall\tau\in\Z^d,[\Norm(\mu+\tau,\Sigma)]_\T=[\Norm(\mu,\Sigma)]_\T$\label{lemma:shift_torus}.
\end{lemma}

\subsection{Limit of the Diffusion on a Torus}
\paragraph*{Proof}
We make a decomposition in Fourier series of a Gaussian $f(x)=\frac{1}{(\sigma\sqrt{2\pi})^d}e^{-\frac{x^2}{2\sigma^2}}$ with frequencies that are multiples of the torus size

\begin{align*}
    \hat{f_\T}(n) & =\int_{\T}f_\T(x)e^{-2\pi i n\cdot x}dx \text{ with } n\in\N^d                                                                                     \\
                  & =\int_{\T}\sum_{\tau\in\Z^d}f(x+\tau)e^{-2\pi i n\cdot x}dx                                                                                        \\
                  & =\sum_{\tau\in\Z^d}\int_{\T}f(x+\tau)e^{-2\pi i n\cdot (x+\tau)}dx                                                                                 \\
                  & =\int_{\R^d}f(x)e^{-2\pi i n\cdot x}dx                                                                                                             \\
                  & =\int_{\R^d}\frac{1}{(\sigma\sqrt{2\pi})^d}e^{-\frac{x^2}{2\sigma^2}}e^{-2\pi i n\cdot x}dx                                                        \\
                  & =\int_{\R^d}\prod_{k=1}^{d}\frac{1}{\sigma\sqrt{2\pi}}\\
                  &e^{-\frac{1}{2\sigma^2}(x_1^2+\cdots+x_d^2)}e^{-2\pi i (n_1\cdot x_1+\cdots+n_d\cdot x_d)}dx \\
                  & =\prod_{k=1}^{d}\int_{\R}\frac{1}{\sigma\sqrt{2\pi}}e^{-\frac{x^2}{2\sigma^2}}e^{-2\pi i n_k\cdot x}dx                                             \\
                  & =\prod_{k=1}^{d}e^{-2\pi\sigma^2 n^2_k}=e^{-2\pi\sigma^2 n^2}                             \\
\end{align*}

Hence $\hat{f_\T}(0_d)=1$ and $\forall n\neq 0_d, \lim_{\sigma\to+\infty}\hat{f_\T}(n)=0$. We conclude that a normal distribution in a torus tends towards a uniform distribution when the variance tends towards infinity.

\subsection{Loss Function}

We first recall equation 18:
\begin{equation*}
    L_{t-1}=\E_q\big[\DKL(q(x_{t-1}|x_t,x_0)||p_\theta(x_{t-1}|x_t))\big]
\end{equation*}

We observe that if we make two Gaussian distributions match in Euclidean space, the distributions will also match in a torus. Also, the implication also holds if we offset the distributions by a multiple of the torus dimension $k\in\Z^d$.
\begin{lemma}
    for all $k\in\Z^d$, the following implication holds.
    \begin{align*}
        (\mu^*,\Sigma^*)=         & \argmin_{\mu,\Sigma}\DKL(\Norm(\mu,\Sigma)||\Norm(\mu'+k,\Sigma'))      \\
        \implies(\mu^*,\Sigma^*)= & \argmin_{\mu,\Sigma}\\ &\DKL([\Norm(\mu,\Sigma)]_\T||[\Norm(\mu'+k,\Sigma')]_\T) \\
        \implies(\mu^*,\Sigma^*)=     & \argmin_{\mu,\Sigma}\DKL([\Norm(\mu,\Sigma)]_\T||[\Norm(\mu',\Sigma')]_\T)
    \end{align*}
\end{lemma}

We can, therefore, choose an arbitrary $k$ for our cost function. The parameters $(\mu, \Sigma)$ that minimize the Kullback Leibler divergence in the Euclidean space $\R^d$ will also minimize it inside of the toric space $\T$.
We can also reformulate the prediction as a trajectory prediction rather than a position prediction by posing $\mu_\theta(x_t,t)=x_t+\epsilon_\theta(x_t,t)$. In this case, for a given parameter $k$ we obtain:

\begin{align*}
    \mathcal{L}^x_k:= & \E_{q,k}\Big[\DKL\big(\Norm(x_0,\tilde{\beta}_t\text{I})||\Norm(\mu_\theta(x_t,t)+k,\sigma^2_t\text{I})\big)\Big] \\
    =                 & \E_{q,k}\big[\frac{1}{2\sigma^2_t}||x_0-\mu_\theta(x_t,t)-k||^2\big]+C                                            \\
    =                 & \E_{q,k}\big[\frac{1}{2\sigma^2_t}||x_0-x_t-k-\epsilon_\theta(x_t,t)||^2\big]+C                                   \\
\end{align*}


We can define the final loss by the shortest trajectory for the displacement of an atom by slightly simplifying the expression. Indeed, the weighting $\frac{1}{2\sigma^2_t}$ can be ignored as it is done in existing work \cite{NEURIPS2020_4c5bcfec}. Moreover, the term $C$ only depends on the variance scheduler $\beta_t$ instead of learned parameters. As such, we can also ignore it.

\begin{equation*}
    \mathcal{L}^x:=\E_{q,k^*}\big[||x_0-x_t-k^*-\epsilon_\theta(x_t,t)||^2]
\end{equation*}
with
\begin{equation*}
    k^*=\argmin_{k\in\Z^d}||x_0-x_t-k||
\end{equation*}

\subsection{Ablation Study}

Here we analyze the importance of the components of our model. The goal is to compare our choice of architecture and assess the relevance of our proposed loss functions. To this end, we conducted three additional experiments with different variants of GNN architectures. The first experiment consists of using equation 9c instead of equation 11 to decode material lattice. This experiment compares the performance of our model when the material is decoded from the noisy cell or from a constant cubic cell (represented as an identity matrix). 

The second experiment evaluates the GNN encoder (equation 9b) using $z^\text{emb}$ instead of $z^\text{enc}$ in equation 9c. This aims to test the relevance of the encoder network before generating the new material geometry. The third experiment is conducted without temporal embedding in equation 8. We modify Equation 9a to not consider the step parameters $t$ as input. We only use a simple embedding depending on the atomic number and change $\mathcal{L}^\rho$ and $\mathcal{L}^x$ to challenge our proposed loss function. To evaluate the relevance of the lattice vector space, we redefine the diffusion process on the lattice by taking $\tilde{\mu}^\rho_t(\rho_t,\rho_0)=\frac{\sqrt{\bar{\alpha}_{t-1}}\beta_t}{1-\bar{\alpha}_t}\rho_0+\frac{\sqrt{\alpha_t}(1-\bar{\alpha}_{t-1})}{1-\bar{\alpha}_t}\rho_t$ instead of $\tilde{\mu}_t(x^\rho_t,x^\rho_0)=\exp\pi\Big(\frac{\sqrt{\bar{\alpha}_{t-1}}\beta_t}{1-\bar{\alpha}_t}x^\rho_0+\frac{\sqrt{\alpha_t}(1-\bar{\alpha}_{t-1})}{1-\bar{\alpha}_t}x^\rho_t\Big)$. The objective of this experiment is to evaluate the relevance of the diffusion process in the lattice vector space. Additionally, we modified $\mathcal{L}^x$ to remove the shortest path constraint. So we used a atomic position loss such that $\mathcal{L}^x:=\mathbb{E}_q\big[||x_0-x_t-\epsilon_\theta(x_t,t)||^2\big]$. In these settings, the loss on atomic position is not in a torus anymore and the atoms should travel without crossing the border of the cell. All the hyperparameters remain the same as shown in Table 1 (in the paper) except the epochs number which is set to 128 epochs. All Table \ref{tab:ablation_study} reports the different results.

\begin{table*}[h]
    \centering
    \begin{tabular}{c|cccc}
        setting & validity & EMD density & EMD energy & FAD \\
        \hline
        proposed model & 99.17\% & 0.289 & 0.205 & 1.602 \\
        from noisy cell instead of cubic$^{(a)}$ & 15.23\% & 223.327 & 1.687 & 48.493 \\
        w/o temporal embedding$^{(b)}$ & 98.78\% & 0.342 & 0.217 & 1.755 \\
        w/o encoder$^{(c)}$ & 11.22\% & 57723.777 & 1.652 & 47.094 \\
        w/o lattice vector space$^{(d)}$ & 99.76\% & 0.392 & 0.277 & 2.701 \\
        w/o shorted path on cell$^{(e)}$ & 97.75\% & 0.292 & 0.206 & 1.691 \\
    \end{tabular}
    \caption{Ablation Study}
    \label{tab:ablation_study}
\end{table*}

In Table 1, we can see that the decoding of the structures from the noisy lattice leads to extremely unstable behaviours. Indeed, configurations (a) and (c) are both decoded from the noisy structure instead of decoded from a cubic cell. Our proposed GNN architecture performs better with temporal embedding than without temporal information (b). However, we notice that the energy of the generated structure without temporal embedding is close to the energy of the proposed model, but the density and validity are lower. Regarding the results of the modified loss function, the performance of the diffusion model clearly dropped  without using the lattice vector space. Indeed, we can see that the density is affected which may reflect instability. This result is expected because adding noise directly in the cell is not a stable operation on the elements of the $\GL$ groups. Moreover, discontinuous action on the cell, such as reflection, is possible. This deformation can be problematic because discontinuities in the diffusion process may lead to instability. Experiments suggest that settings (d) are less stable than our proposed model. Finally, the experiment performed without periodic constraints on the atomic position results in a score closely related to our proposed model for the density and energy of the generated structures. However, the FAD denotes a lower performance and the validity metric suggests more atomic collision than our proposed model. We believe that the atomic position is less precise without taking the periodicity of the space into account. Finally, we notice that, besides their different behaviour, the loss functions have a positive impact on the generations of atomic position and lattice shape.

\subsection{Property Distribution}

\begin{figure}
\centering
\includesvg[width=0.48\textwidth]{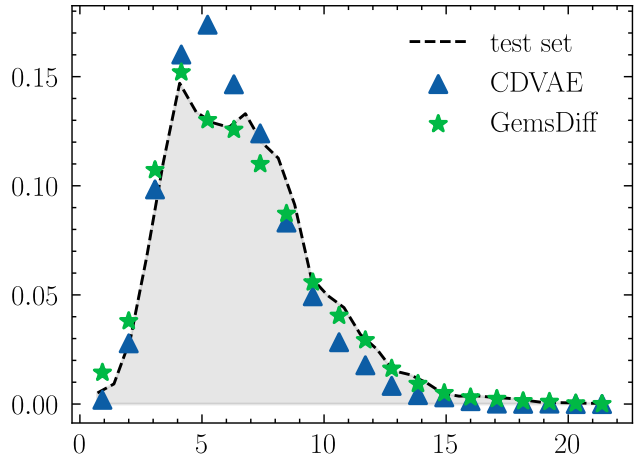}
\caption{Density ($g.cm^{-3}$) distribution over the generated structure set}\label{fig:density}
\end{figure}

\begin{figure}
\centering
\includesvg[width=0.48\textwidth]{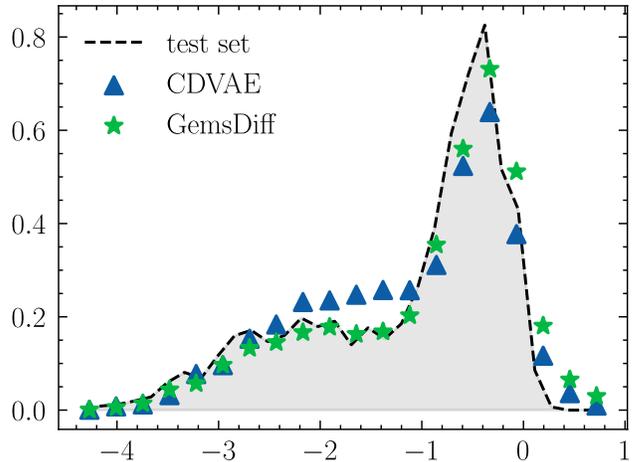}
\caption{Energy ($eV.\text{atom}^{-1}$) distribution over the generated structure set}\label{fig:energy}
\end{figure}

Here we deeply analyze the distribution of properties. Figure \ref{fig:density} and \ref{fig:energy} represent the distributions of the energy and density. We include a set of structures generated by CDVAE \cite{xie2021crystal} and GemsDiff.  The real distribution of each property is also provided for reference. We can see that Both CDVAE and GemsDiff are closely correlated with a real distribution of the features. However, we can observe that GemsDiff remains much closer to the real distribution data than CDVAE. Indeed, there is a gap between CDVAE and the real distribution around $5 g.cm^{-3}$ for the density of the structures, and  between $-2.5$ and $-1 eV.\text{atom}^{-1}$ for the energy. 


\end{document}